# One million year old groundwater in the Sahara revealed by krypton-81 and chlorine-36

N. C. Sturchio,[1] X. Du,[2,3] R. Purtschert,[4] B. E. Lehmann,[4] M. Sultan,[5] L. J. Patterson,[1] Z.-T. Lu,[2] P. Müller,[2] T. Bigler,[4] K. Bailey,[2] T. P. O'Connor,[2] L. Young,[6] R. Lorenzo,[4] R. Becker,[5] Z. El Alfy,[7] B. El Kaliouby,[8] Y. Dawood,[8] and A. M. A. Abdallah[8]

[1]*Department of Earth and Environmental Sciences, University of Illinois at Chicago, Chicago, USA*
[2]*Physics Division, Argonne National Laboratory, Argonne, USA*
[3]*Physics Department, Northwestern University, Evanston, USA*
[4]*Institute of Physics, University of Bern, Bern, Switzerland*
[5]*Department of Geology, State University of New York at Buffalo, Buffalo, USA*
[6]*Chemistry Division, Argonne National Laboratory, Argonne, USA*
[7]*Egyptian Geological Survey and Mining Authority, Cairo, Egypt*
[8]*Department of Geology, Ain Shams University, Cairo, Egypt*

**Abstract.** Measurements of $^{81}$Kr/Kr in deep groundwater from the Nubian Aquifer (Egypt) were performed by a new laser-based atom-counting method. $^{81}$Kr ages range from ~2×10$^5$ to ~1×10$^6$ yr, correlate with $^{36}$Cl/Cl ratios, and are consistent with lateral flow of groundwater from a recharge area near the Uweinat Uplift in SW Egypt. Low $\delta^2$H values of the $^{81}$Kr-dated groundwater reveal a recurrent Atlantic moisture source during Pleistocene pluvial periods. These results indicate that the $^{81}$Kr method for dating old groundwater is robust and such measurements can now be applied to a wide range of hydrologic problems.

The age of groundwater is one of the most elusive geologic parameters to quantify, despite its crucial significance for water resources, waste management, subsurface reactive transport, and paleoclimate. Groundwater age is usually defined as the mean subsurface residence time following isolation from the atmosphere, and it can be estimated either from Darcy's Law (based upon hydraulic conductivity and gradient) or from measurements of time-dependent abundances of natural isotopic tracers. The only available quantitative method for dating old (5×10$^4$–10$^6$ yr) groundwater involves measurements of cosmogenic $^{36}$Cl ($t_{1/2}$ = 3.01×10$^5$ yr) [Phillips, 2000]. The $^{36}$Cl method is complicated by variations of the initial $^{36}$Cl activity and by subsurface input of both stable chloride (Cl) and nucleogenic $^{36}$Cl [Park et al., 2002]. A more optimal method for dating old groundwater is based on cosmogenic $^{81}$Kr ($t_{1/2}$ = 2.29×10$^5$ yr), which has a more constant initial activity than $^{36}$Cl and negligible subsurface source [Loosli and Oeschger, 1969], but has been heretofore almost impossible to measure. Many groundwater aquifers do not meet the restrictive criteria for application of the $^{36}$Cl method, such as those containing saline waters and brines and those consisting of fractured igneous and metamorphic rocks, and therefore the $^{36}$Cl method cannot be applied to dating waters in these aquifers. The capability of making routine $^{81}$Kr measurements in groundwater and other environmental samples (e.g., glacial ice and hydrothermal fluid discharges) is needed, and in this paper we present results demonstrating a significant advance toward the routine application of this method in hydrologic studies.

$^{81}$Kr ($t_{1/2} = 2.29 \times 10^5$ yr) is produced in the upper atmosphere by cosmic-ray induced spallation and neutron activation of stable Kr isotopes [Loosli and Oeschger, 1969]. As a result of its long lifetime and chemical inertness in the atmosphere, $^{81}$Kr is expected to have a constant and well-constrained atmospheric source with negligible subsurface sources or sinks other than radioactive decay. The $^{81}$Kr/Kr ratios of samples extracted from air before and after the development of the nuclear industry have been measured and were found to be identical within the measurement error of ±8% [Du et al., 2003]. The difficulty in analyzing $^{81}$Kr, owing to its low isotopic abundance ($^{81}$Kr/Kr ~ $10^{-12}$) and low solubility of Kr in water (e.g. one liter of surface water equilibrated with air at 20°C and 1 atm contains only $7 \times 10^{-5}$ cm$^3$ STP of Kr and $2 \times 10^3$ $^{81}$Kr atoms), has been its main drawback. Two earlier methods of analyzing $^{81}$Kr were applied to very old groundwater. The first used resonance ionization mass spectrometry (RIMS) [Thonnard et al., 1987] after two steps of isotope enrichment to measure $^{81}$Kr extracted from 50 kg of water from the Milk River Aquifer, Canada [Lehmann et al., 1991]. The second, following extraction of Kr from 16-ton samples of groundwater from the Great Artesian Basin, Australia, used accelerator mass spectrometry (AMS) measurements of $^{81}$Kr with a high-end cyclotron [Collon et al., 2000]. Both methods were successful in their proof-of-principle measurements; however, their routine use for $^{81}$Kr measurement is hindered by the complex enrichment procedures required for RIMS and the limited accessibility of large accelerator facilities for AMS.

Recent developments in the Atom Trap Trace Analysis (ATTA) method, based upon laser manipulation of neutral Kr atoms [Chen et al., 1999; Du et al., 2003], have enabled the analysis of $^{81}$Kr in Kr gas extracted directly from ~2-ton groundwater samples using a tabletop apparatus, thus providing a more practical approach for dating very old groundwater. This paper demonstrates the first use of ATTA for groundwater dating by the $^{81}$Kr method. The Nubian Aquifer of the Western Desert of Egypt [Thorweihe, 1990] was selected for study because of its relatively simple geology, potential for containing very old groundwater, and favorable characteristics for comparison of the $^{81}$Kr and $^{36}$Cl methods. Until this work, no definitive evidence has been produced for either a gradient in groundwater age within the Nubian Aquifer or an upper limit to that age.

Samples of Nubian Aquifer groundwater from wells in the Western Desert of Egypt were measured for $^{81}$Kr, $^{36}$Cl, and other chemical and isotopic constituents (Fig. 1). For $^{81}$Kr dating, dissolved gas was extracted from several tons of water at six field sites using a vacuum-stripping method [Collon et al., 2000]. At the University of Bern, Kr was separated from the gas samples by molecular sieve absorption and gas chromatographic methods; low-level counting was used to confirm that the abundances of $^{85}$Kr ($t_{1/2} = 10.8$ yr) in these samples were indeed low ($^{85}$Kr/Kr <3% of modern air) as expected for old groundwaters, verifying minimal air contamination during sampling; and, for normalization in ATTA analyses, a calibrated amount of $^{85}$Kr was mixed with each Kr sample. The $^{81}$Kr/$^{85}$Kr ratios in the spiked Kr samples were then measured using ATTA at Argonne National Laboratory [Du et al., 2003]. Chloride concentrations in water samples were measured by ion chromatography and Cl was then precipitated as AgCl for measurements of $^{36}Cl$ that were done by accelerator mass spectrometry at the PRIME Lab of Purdue University.

Results of the groundwater analyses are shown in Table 1. $^{81}Kr$ is expressed in terms of the air-normalized ratio, $R/R_{air} = [^{81}Kr/Kr]_{sample}/[^{81}Kr/Kr]_{air}$, where $R_{air}$ is the modern atmospheric ratio $[^{81}Kr/Kr]_{air}=1.10(\pm0.05)\times10^{-12}$ measured by ATTA [Du et al., 2003]. $R/R_{air}$ values range from 4.8% to 52.6%. Using the simple expression for radioactive decay, and the $^{81}Kr$ decay constant $\lambda_{Kr} = 3.03(\pm0.14)\times10^{-6}$ yr$^{-1}$, the age $t_{Kr}$ of a groundwater sample is given by

$$t_{Kr} = -1/\lambda_{Kr} \ln (R/R_{air}). \qquad (1)$$

The range of ages thus derived is $0.2–1.0\times10^6$ yr; these are *apparent* ages because the extent of mixing within the sampled wells is unknown. Ages increase progressively along flow vectors predicted by numerical hydrodynamic models [Brinkmann et al., 1987], verifying distant lateral flow of deep groundwater toward the northeast from a recharge area southwest of Dakhla. The general correspondence of $^{81}Kr$ age with predicted hydrodynamic age indicates negligible input of $^{81}Kr$ from subsurface sources, in agreement with conclusions reached from $^{81}Kr$ measurements of groundwater from the Milk River Aquifer in Canada [Lehmann et al., 1991] and the Great Artesian Basin in Australia [Collon et al., 2000]. The estimated diffusive loss of $^{81}Kr$ to aquitards was limited to ≤20% for the Great Artesian Basin [Lehmann et al., 2003], and is expected to be even lower within the Nubian Aquifer because of the lower porosity and smaller proportion of aquitard formations.

$^{81}Kr$ ages increase progressively with distance to the north and east of Dakhla; their spatial distribution indicates relatively high flow velocities (~2 m/yr) from Dakhla toward Farafra, and low velocities (~0.2 m/yr) from Dakhla toward Kharga and Baris and from Farafra to Bahariya. These observations are consistent with the areal distribution of hydraulically conductive sandstone within the aquifer and they provide support to some of the existing hydrodynamic models [Brinkmann et al., 1987; Hesse et al., 1987]. Southwestward extrapolation of the ~2 m/yr flow rate inferred from the difference in $^{81}Kr$ ages for Dakhla and Farafra is consistent with recharge in the area of the Uweinat Uplift near the Egypt-Sudan border (Fig. 1). In this area, the Nubian sandstone is exposed (or buried beneath sand sheets or dunes) at elevations between 200 and 600 m above sea level over a wide area, forming a broad catchment for recharge of the Nubian Aquifer.

There is a good correlation between $^{36}Cl/Cl$ ratios and $^{81}Kr$ ages (Fig. 2), in samples for which both measurements were made, providing a firm basis for calibrating apparent $^{36}Cl$ ages to an apparent value of $[^{36}Cl/Cl]_{initial}$ for the Nubian Aquifer. A best-fit curve through the data using the function

$[^{36}Cl/Cl]_{measured}$
$= [^{36}Cl/Cl]_{seq}+ ([^{36}Cl/Cl]_{initial} - [^{36}Cl/Cl]_{seq}) \times \exp(-t_{Kr}\times\lambda_{Cl}) \qquad (2)$

where $t_{Kr}$ is the [81]Kr-age in years, $\lambda_{Cl} = 2.30(\pm0.02)\times10^{-6}$ yr$^{-1}$ is the [36]Cl decay constant, and $[^{36}Cl/Cl]_{seq} = 8(\pm3)\times10^{-15}$ is the secular equilibrium in-situ production value for the aquifer [Phillips, 2000] (assuming constant Cl concentration in water) gives $[^{36}Cl/Cl]_{initial} = 131(\pm11)\times10^{-15}$, which agrees well with the initial value of $125(\pm10)\times10^{-15}$ estimated for the Great Artesian Basin [Love et al., 2000]. [81]Kr and [36]Cl ages for four out of six samples agree within 1σ, but two samples exhibit significant age discordance. Sample El Zayat 12 (Cl = 59 mg/L) has a [36]Cl age of $2.8\times10^5$ yr [Eq. (2)] that is significantly younger than its [81]Kr age of $3.9\times10^5$ yr, indicating that it may have had a higher initial value $[^{36}Cl/Cl]_{initial}$ or it may have experienced subsurface input of nucleogenic [36]Cl. Sample Sherka-36 (Cl = 95 mg/L) has a [36]Cl age of $1.5\times10^6$ yr [Eq. (2)] that is significantly older than its [81]Kr age of $6.8\times10^5$ yr. Here the diagreement may be explained by the addition of subsurface Cl having low [36]Cl/Cl; such Cl could be obtained either by dissolution of Cl-bearing minerals or by diffusion from stagnant, saline pore fluids in aquitards. Stable Cl isotope ratios and Cl concentrations for these and other Nubian Aquifer groundwater samples measured for [36]Cl support such explanations of the observed [81]Kr-[36]Cl discordance [Patterson, 2003; Patterson et al., 2003]. These results demonstrate that the integrity of the [36]Cl "clock" is vulnerable to additions of Cl at any point along a flowpath, whereas, in contrast, the potential for open-system behavior of Kr is minimal after water has reached the water table.

Stable isotope ratios of hydrogen and oxygen in the Nubian Aquifer (and equivalent) groundwaters from across Saharan North Africa indicate a clearly defined continental isotope effect involving precipitation from air masses having an Atlantic moisture source; δ[2]H values decrease gradually from west to east with a geographic pattern resembling that observed in modern European groundwaters [Sonntag et al., 1978]. In the deep Nubian Aquifer samples, nearly constant δ[2]H values (from −82 to −79) as a function of [81]Kr age reveal that this Atlantic moisture source was recurrent during all major pre-Holocene pluvial periods throughout the past $1\times10^6$ yr. In contrast, the isotopic composition of modern precipitation in the study area, which mostly delivers moisture from the Mediterranean, is much less depleted in [2]H (δ[2]H ranges from −28 to +14) [IAEA, 1990; Bakri et al., 1992].

Inferences regarding the recharge history of the Nubian Aquifer, based on the chronology developed from [81]Kr and [36]Cl measurements, corroborate available chronologies for other terrestrial and marine paleoclimate indicators in the region [Rossignol-Strick, 1983; Szabo et al., 1989; Szabo et al., 1995; Sultan et al., 1997; Crombie et al., 1997]. The picture revealed by the Nubian Aquifer groundwater archive is a diffuse reflection of the climate history of Northeast Africa for the past ~$1\times10^6$ yr. Investigations of other such archives may yield useful insights into the hydrology of large groundwater basins (both terrestrial and marine), the climatic evolution of continental interiors, and groundwater-climate linkages throughout the Late Quaternary. This study verifies that the [81]Kr method is a robust tool for dating groundwater up to about one million years old. Application of this method to a wide range of hydrologic problems, even those involving saline waters and brines, appears now to be feasible on a routine basis.

**Acknowledgments.** This work was supported mainly by the U. S. National Science Foundation (Grant EAR-0126297) and the U. S. Department of Energy, Office of Nuclear Physics (Contract W-31-109-Eng-38). Partial support was provided by NASA's Land-Cover Land-Use Changes program (Grant NAG5-12409); the U. S. Department of Energy, Office of Basic Energy Sciences (Contract W-31-109-Eng-38); the Swiss National Science Foundation; and the University of Bern.

Mailing address: L.J. Patterson, N.C. Sturchio, *Department of Earth and Environmental Sciences, University of Illinois at Chicago, Chicago, Illinois 60637, USA. (sturchio@uic.edu)*

K. Bailey, X. Du, Z.-T. Lu, P. Mueller, T. P. O'Connor, L. Young, *Physics Division, Argonne National Laboratory, Argonne, IL 60439, USA. (lu@anl.gov)*

T. Bigler, B. Lehmann, R. Lorenzo, R. Purtschert, *Climate and Environmental Physics, Physics Institute, University of Bern, Sidlerstr. 5, 3012 Bern, Switzerland. (purtschert@climate.unibe.ch)*

R. Becker and M. Sultan,*Department of Geology, 876 Natural Sciences Complex, Buffalo, NY 14260, USA. (misultan@geology.buffalo.edu)*

Z. El Alfy, *Egyptian Geological Survey and Mining Authority, 3. salah salem st., Abbassiya, Cairo, Egypt.*

B. El Kaliouby, Y. Dawood, and A.M.A. Abdallah, Department of Geology, Ain Shams University, Abbassia 11566, Cairo, Egypt.

STURCHIO ET AL.: NUBIAN AQUIFER AGE

**Fig. 1.** Map showing sample locations (red circles) in relation to oasis areas (shaded green), Precambrian basement outcrops (patterned), and other regional features. Groundwater flow in Nubian Aquifer is toward northeast.

**Fig. 2.** $^{36}$Cl/Cl ($\times 10^{-15}$) vs. $^{81}$Kr age for Nubian Aquifer groundwater samples ($\pm 1\sigma$ error bars), showing best-fit exponential decay curve of $^{36}$Cl. Intercept on y-axis represents $[^{36}Cl/Cl]_{initial}$, the initial $^{36}$Cl/Cl ratio of groundwater, which is $131(\pm 11) \times 10^{-15}$ when $[^{36}Cl/Cl]_{seq} = 8(\pm 3) \times 10^{-15}$ is assumed for the secular equilibrium value of $^{36}$Cl/Cl in the sandstone.

**Table 1.** $^{81}$Kr and $^{36}$Cl data for Nubian Aquifer groundwaters.

| Well Name* | Depth (m) | Cl$^-$ (mg/L) | $^{36}$Cl/Cl ($\times 10^{-15}$) | $^{81}$Kr/Kr (R/R$_{air}$)% | $^{81}$Kr-Age $t_{Kr}$ ($\times 10^5$ yr) | $\delta^2$H (‰) |
|---|---|---|---|---|---|---|
| Gum Horia | 1200 | 20 | 76.5 (±3.4) | 52.6 (±6.1) | 2.1 (±0.4) | −81 |
| Farafra 6 | 800 | 24 | 65 (±3) | 36.5 (±4.2) | 3.3 (±0.4) | −79 |
| Bauti 1 | 1200 | 52 | 20.2 (±1.6) | 4.8 (±3.8) | 10 (+6 / -2) | −81 |
| El Zayat 12 | 720 | 59 | 72.7 (±2.7) | 30.6 (±3.6) | 3.9 (±0.4) | −82 |
| Baris (Aden) | 600 | 92 | 45.6 (±2.1) | 22.8 (±3.0) | 4.9 (±0.5) | −81 |
| Sherka 36 | 750 | 95 | 12.2 (±5.1) | 12.8 (±3.0) | 6.8 (±0.8) | −82 |

*well locations shown in Fig. 1.

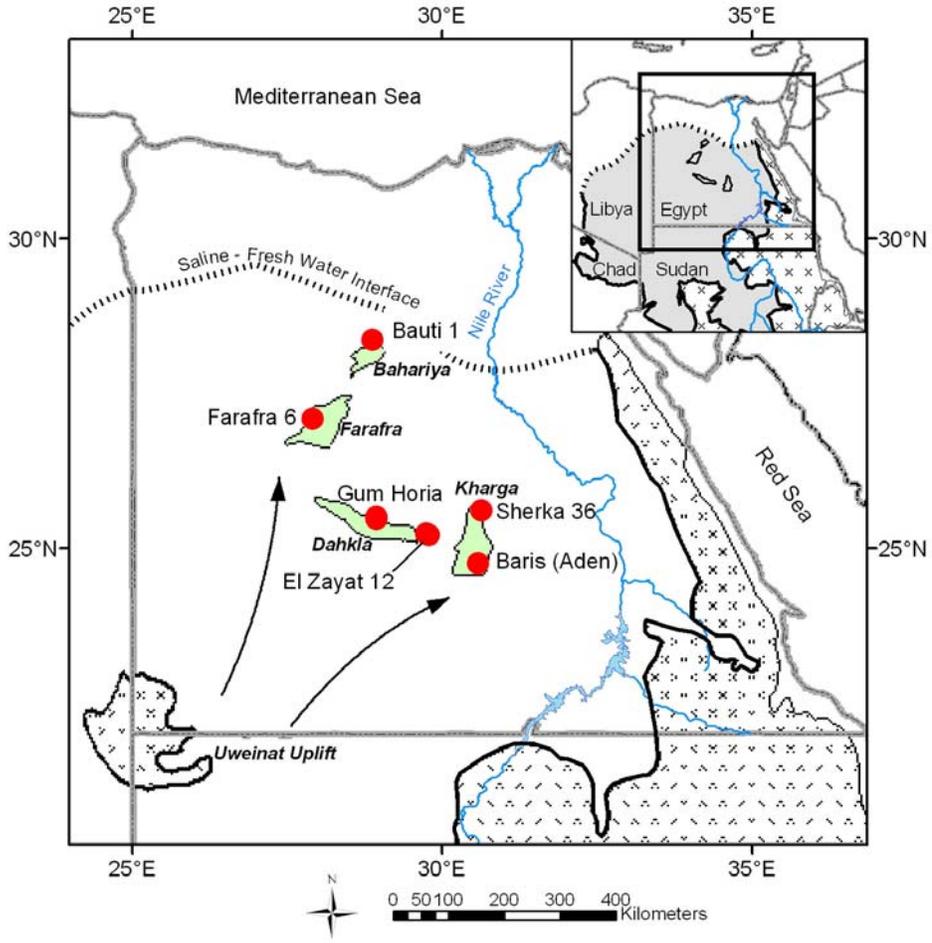

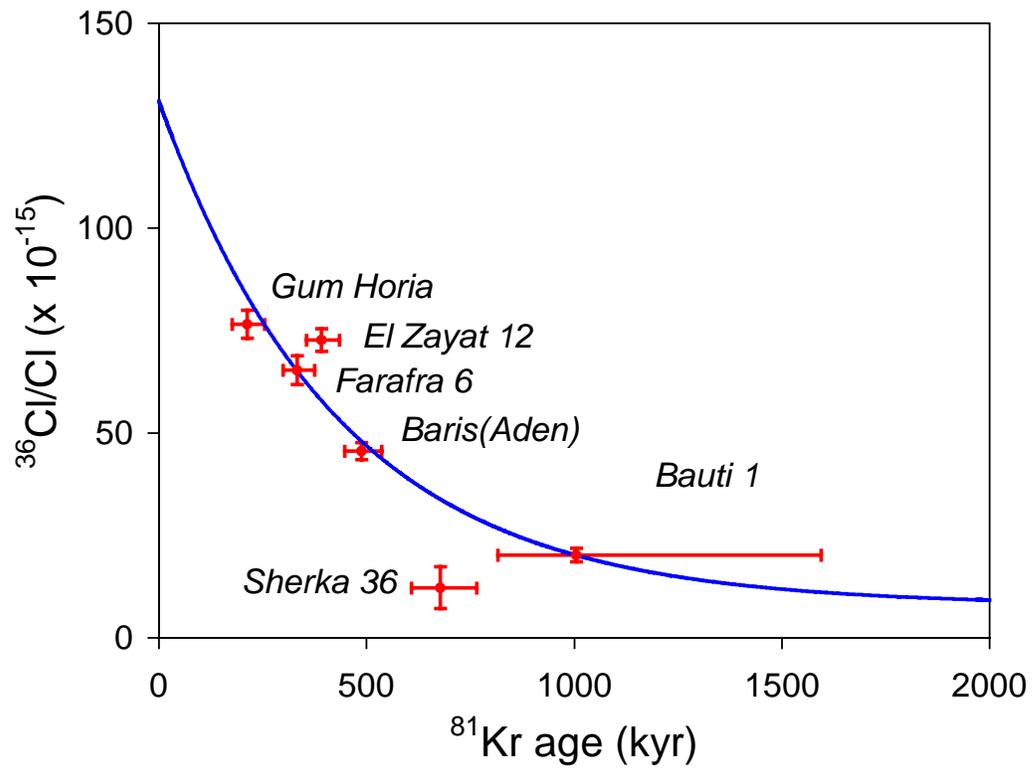